\documentclass{UniDraft}

\usepackage[
    doi=false,
    isbn=false,
    url=false,
    eprint=true,
    giveninits=true,
    date=year,
    sorting = none,
    style = numeric-comp,
    maxbibnames=99
]{biblatex}
\newcommand{\ii}{\mathrm{i}}
\newcommand{\ee}{\mathrm{e}}

\title{\bfseries \large Observation of Universal Spectral Moments and the Dynamic Dispersive-to-Proliferative Transition}

\author[1]{Jia-Xin Zhong}
\author[2]{Chang Shu}
\author[2]{Nan Cheng}
\author[1]{Jee Woo Kim}
\author[2]{Kai Zhang}
\author[2,*]{Kai Sun}
\author[1,$\dagger$]{Yun Jing}

\affil[1]{Graduate Program in Acoustics, The Pennsylvania State University, University Park, PA 16802, USA}
\affil[2]{Department of Physics, University of Michigan, Ann Arbor, Michigan 48109, United States}

\date{\vspace{-1em}\small 
    \textsuperscript{*}sunkai@umich.edu;
    \textsuperscript{$\dagger$}yqj5201@psu.edu
}

\titleformat{\section}         
  {\normalfont\Large\bfseries} 
  {}                           
  {0pt}                        
  {}                           

\titleformat{\subsection}      
  {\normalfont\large\bfseries} 
  {}                           
  {0pt}                        
  {}
\titlespacing*{\section}{0pt}{3.5ex plus 1ex minus .2ex}{2.3ex plus .2ex}
\titlespacing*{\subsection}{0pt}{3.25ex plus 1ex minus .2ex}{1.5ex plus .2ex}

\usepackage{abstract}

\begin{document}


\pdfbookmark[1]{Title}{title} 

\maketitle



\begin{abstract}

In non-Hermitian systems, spectra can be maximally boundary-sensitive, yet bulk physics need not be.
Here we experimentally show that spectral moments provide boundary-robust bulk observables in finite non-Hermitian lattices, even when the spectra undergo dramatic geometry-dependent reshaping due to the skin effect.
Using a unified acoustic platform with full spectral reconstruction and time-domain access, we probe one-, two- and three-dimensional lattices and demonstrate that spectral moments remain nearly invariant across distinct boundary geometries while the corresponding complex spectra differ strongly.
To connect the thermodynamic theorem to realistic finite systems, we develop a loop-counting theory that identifies the physical origin of finite-size deviations in terms of missing boundary loops, quantitatively captures the corrections, and predicts a scaling law, which we verify experimentally.
Beyond acoustic spectroscopy, we reveal a counterintuitive dynamical consequence of moment invariance: a dispersive-to-proliferative bulk transition governed by bulk moment structure rather than spectral boundary sensitivity.
As a result, local bulk dynamics can remain stable (dispersive) even in a $\mathcal{PT}$-broken spectral regime, challenging the conventional expectation that $\mathcal{PT}$ breaking necessarily implies feedback-induced dynamical instability (proliferation) through exponentially amplifying spectral components. 
These results establish spectral moments as practical bulk descriptors for finite non-Hermitian matter and open a route to extracting and controlling intrinsic bulk behavior in realistic wave-based non-Hermitian devices.

\end{abstract}

\thispagestyle{firststyle}

\newpage

\section{Introduction}
Non-Hermitian physics provides a general framework for wave and quantum systems that exchange energy with their environment, enabling phenomena with no Hermitian counterpart~\cite{Ashida2020NonHermitianPhysics,ZhenBo2015,Longhi2015SR,Zhou2018,McDonaldPRX2018,Cerjan2019NP,Vincenzo2020,Sebastian2020Science, Bergholtz2021ExceptionalTopologyNonHermitian, MaGC2022Nature,Huang2024NRP}. 
Among them, the non-Hermitian skin effect (NHSE) has emerged as a defining paradigm in lattice systems: as a consequence of non-Hermitian bulk structure, often but not necessarily associated with non-reciprocal couplings, an extensive fraction of eigenstates can accumulate near boundaries, accompanied by an anomalous sensitivity of spectra and eigenstates to boundary conditions~\cite{LeeModel2016,Yao2018,Torres2018,Kunst2018BiorthogonalBulkboundaryCorrespondence,Xiong2018, Murakami2019PRL, Lee2019AnatomySkinModes,Ashvin2019PRL, Kai2020, Okuma2020TopologicalOriginNonHermitian, Helbig2020GeneralizedBulkBoundary, Xiao2020NonHermitianBulkBoundary, Weidemann2020TopologicalFunnelingLight,XuePeng2020,Ghatak2020ObservationNonHermitianTopology, Wang2021DetectingNonBlochTopological, Ding2022NonHermitianTopologyExceptionalpoint, Zhang2022ReviewNonHermitianSkin, Kai2022NC, Okuma2023NonHermitianTopologicalPhenomena, Lin2023TopologicalNonHermitianSkin,SBZhang2023PRL,AmoebaPRX2024,KaiASE2025}. 
A direct consequence is an extreme sensitivity of the complex spectrum to boundary conditions and boundary geometry~\cite{Budich2020NonHermitianTopologicalSensors,Clerk2020NC,Li2020CriticalNHSE,WTXue2021,CXGuo2021,Micallo2023,ultrasensitivity2025}. 
This boundary ultrasensitivity raises a fundamental question for theory, experiments, and applications alike: if the spectrum is strongly reshaped by boundary conditions, while wave propagation deep in the bulk should be determined solely by the bulk environment and remain insensitive to the boundary, what physical quantity should be used to characterize intrinsic bulk propagation?

Recent theory has identified a striking answer. 
Intuitively, a spectral moment is a compact descriptor of the eigenvalue distribution, obtained by averaging powers of the eigenvalues, much like moments of a probability distribution characterize its shape.
The universal spectral moment theorem shows that, despite dramatic spectral reshaping, the normalized spectral moments (equivalently, normalized traces of powers of the Hamiltonian) converge in the thermodynamic limit to quantities determined only by bulk hopping structure, independent of boundary geometry~\cite{Cheng2024UniversalSpectralMoment}. 
This result reframes the NHSE from a purely spectral pathology into a setting where bulk information remains encoded in carefully chosen observables. 
Moreover, the same framework predicts a boundary-agnostic dynamical classification---including a dispersive-to-proliferative transition governed by the bulk moment structure---suggesting that short-time local dynamics can diagnose intrinsic bulk stability even when the global spectrum is strongly boundary sensitive.

Classical wave metamaterials have become a central experimental route to non-Hermitian physics, with photonic, mechanical, electrical and acoustic platforms allowing controlled implementations of gain, loss, coupling and boundary conditions \cite{Zhang2021AcousticNonHermitianSkin, Liu2022ExperimentalRealizationWeyl, Chen2024RobustTemporalAdiabatic, Chen2025DirectMeasurementTopological, Tong2025ObservationFloquetBlochBraids, Zhong2025HigherorderSkinEffect, Wu2025ObservationDislocationNonHermitian, Zhong2026ObservationDislocationBound,Zhong2025ExperimentallyProbingNonHermitian, Zhong2026ObservationErraticNonHermitian, Wang2025AndersonTransitionComplex, Weidemann2021CoexistenceDynamicalDelocalization}.
Yet translating the spectral-moment framework into experiment remains nontrivial.
Most importantly, it requires a unified spectral-and-dynamical platform: the same device must provide programmable non-Hermitian couplings, reconfigurable boundary geometries, full complex-spectrum reconstruction, and local time-domain wave-packet tracking.
Such combined access is essential for testing not only the boundary robustness of spectral moments, but also the predicted dispersive-to-proliferative transition, where local bulk stability can be decoupled from conventional expectations based on boundary-sensitive spectra in a uniform-loss background.
A second challenge is that the theorem is asymptotic, whereas all experimental realizations are finite-size systems.
The magnitude and scaling of finite-size deviations must therefore be quantified before spectral moments can serve as practical bulk observables.

Here we realize this program using acoustic metamaterials \cite{Zhang2021AcousticNonHermitianSkin, Liu2022ExperimentalRealizationWeyl, Chen2024RobustTemporalAdiabatic, Chen2025DirectMeasurementTopological, Tong2025ObservationFloquetBlochBraids, Zhong2025HigherorderSkinEffect,  Wu2025ObservationDislocationNonHermitian, Zhong2026ObservationDislocationBound,Zhong2025ExperimentallyProbingNonHermitian, Zhong2026ObservationErraticNonHermitian} and experimentally validate the universal spectral moment theorem together with its dynamical implications. 
Building on our unified spectral-and-dynamical acoustic platform for full complex-spectrum reconstruction and time-domain wave-packet measurements on the same active lattice \cite{Zhong2025ExperimentallyProbingNonHermitian, Zhong2026ObservationErraticNonHermitian}, we implement non-Hermitian tight-binding models with reconfigurable boundaries in one, two and three dimensions. 
We show that spectral moments remain nearly invariant across sharply different boundary geometries even when the spectra themselves are drastically reshaped, and we derive a loop-counting theory that yields a universal finite-size scaling law for the residual deviation, which is verified experimentally. 
Beyond acoustic spectroscopy, our time-domain measurements reveal the predicted dispersive-to-proliferative transition and establish that local bulk dynamics can remain dispersive in regimes where the spectrum is $\mathcal{PT}$-broken. 
Together, these results demonstrate a practical route to extracting and controlling intrinsic bulk behavior in finite non-Hermitian wave systems, and open a pathway toward boundary-robust functionalities in realistic non-Hermitian devices.

\section{Results}

\subsection{Spectral moments in finite-size systems}

\begin{figure}[!htbp]
    \centering
    \phantomsection 
    \includegraphics[width=0.99\linewidth]{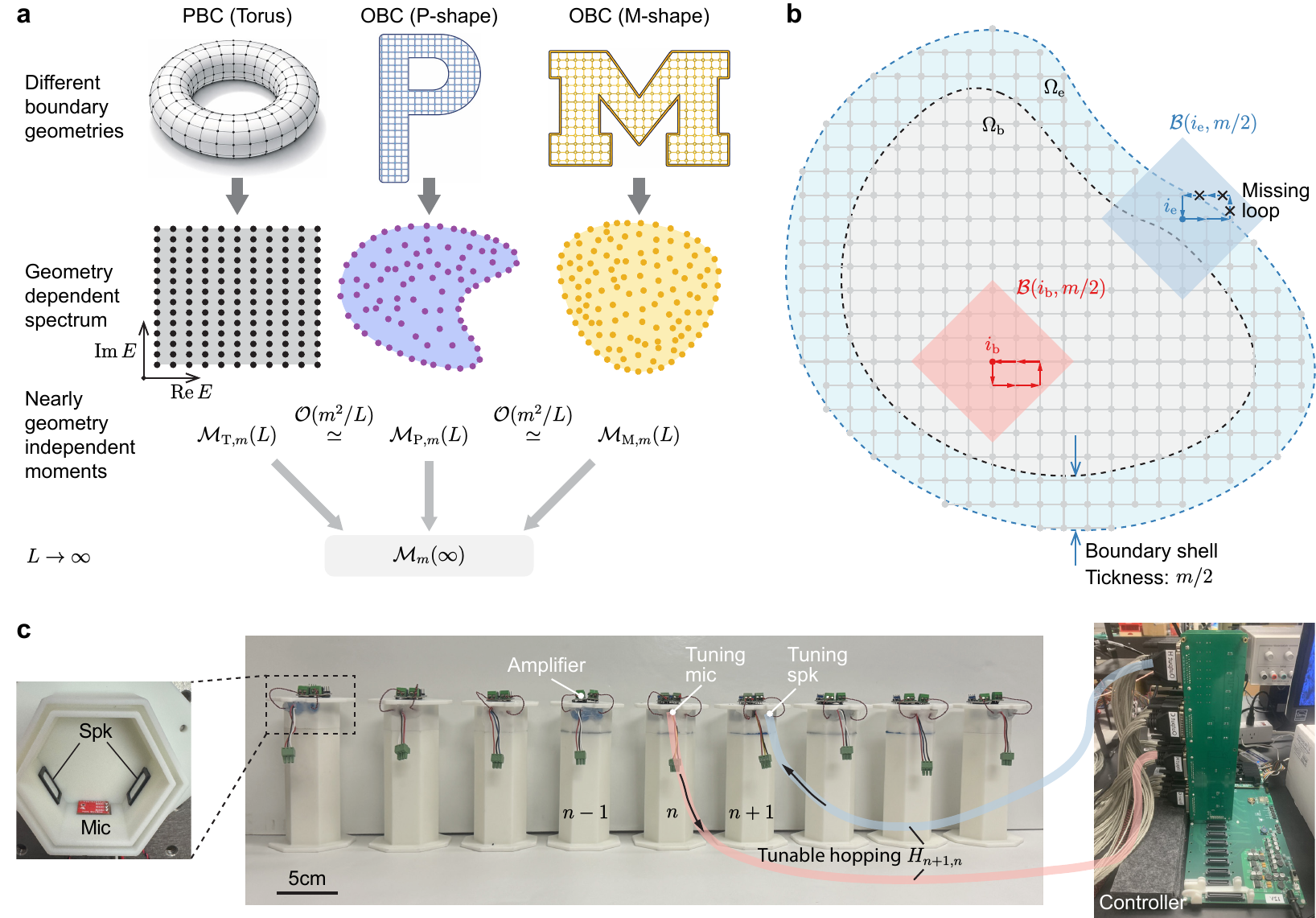}
    \caption{
        \textbf{Boundary-sensitive spectra versus boundary-robust spectral moments, and the acoustic lattice platform.}
        \textbf{a,} For the same bulk non-Hermitian lattice Hamiltonian at finite size $L$, the complex spectrum $\{E_n\}$ reshapes drastically under different boundary geometries (PBC torus versus OBC P- and M-shaped domains), whereas the per-site spectral moments $\mathcal{M}_{\alpha,m}(L)$ are nearly boundary independent. The residual geometry dependence scales as a universal finite-size correction $\mathcal{O}(m^2/L)$, and $\mathcal{M}_{\alpha,m}(L)\to \mathcal{M}_m(\infty)$ as $L\to\infty$.
        \textbf{b,} Loop-counting picture for the finite-size deviation. The dashed curve partitions the domain into a bulk region $\Omega_\mathrm{b}$ and a boundary shell $\Omega_\mathrm{e}$ of thickness $\sim m/2$ (in hop distance). For a bulk rooting site $i_\mathrm{b}\in\Omega_\mathrm{b}$, all length-$m$ closed walks can be sampled within the hop-distance ball $\mathcal{B}(i_\mathrm{b},m/2)$ (red shaded diamond), yielding the fully sampled loop-weight sum $w_m$. For a boundary-shell site $i_\mathrm{e}\in\Omega_\mathrm{e}$, the ball $\mathcal{B}(i_\mathrm{e},m/2)$ (blue shaded diamond) intersects the exterior of the domain, so some length-$m$ loops are missing (crosses), producing the finite-size correction.
        \textbf{c,} Photograph of the experimental platform: a chain of acoustic resonators (interior view shown at left, containing a microphone and two loudspeakers) coupled by electroacoustic feedback implemented with external amplifiers and phase shifters integrated in a multi-channel controller (right). A representative programmable hopping $H_{n+1,n}$ from site $n$ to site $n+1$ is indicated.
    }
    \label{fig:loop_counting}. 
    \addcontentsline{toc}{subsection}{Figure \thefigure}
\end{figure}

Ideally, spectral moments are insensitive to the boundary shape and capture the essential bulk physics of a non-Hermitian lattice system in arbitrary dimensions (Fig.~\ref{fig:loop_counting}a).
However, the spectral moments are boundary-agnostic only in the thermodynamic limit~\cite{Cheng2024UniversalSpectralMoment}.
For a finite-size system of linear size $L$, the spectral moments generally deviate from their bulk (thermodynamic) values.
As we show below, this finite-size deviation exhibits a scaling law with the system size $L$. 
We further elucidate the physical origin of the deviation and derive this scaling law, which will be verified experimentally in our active acoustic platforms (Fig.~\ref{fig:loop_counting}c).

Consider a $d$-dimensional lattice with a tight-binding Hamiltonian $H$, linear size $L$, and total number of sites $N$.
For brevity, we focus on a single-band model, where $H_{ij}$ denotes the hopping amplitude from site $j$ to site $i$ ($i\neq j$).
Without loss of generality, we subtract the background potential and assume $H_{ii}=0$.
The $m$-th spectral moment is given by
\begin{equation}
\Tr H^m=\sum_{i_1,\ldots,i_m} H_{i_1 i_2} H_{i_2 i_3}\cdots H_{i_m i_1}.
\end{equation}
The index sequence $(i_1\to i_2\to \cdots \to i_m\to i_1)$ forms a closed walk (loop) of length $m$ on the lattice.
Denoting a length-$m$ loop by $O$ and its weight by $w(O)$, one may rewrite
$
\Tr H^m=\sum_{|O|=m}w(O).
$
Equivalently, one may reorganize the sum by first choosing the rooting site $i$ and then summing over all length-$m$ loops $O_i$ rooted at $i$:
\begin{equation}\label{eq:bulk_sum}
\Tr H^m=\sum_i \sum_{|O_i|=m}w(O_i).
\end{equation}

In the thermodynamic limit, the lattice is effectively infinite.
For a bulk rooting site $i$, a length-$m$ loop can be sampled entirely within the set of sites whose graph (hop) distance to $i$ is at most $m/2$; see Fig.~\ref{fig:loop_counting}b.
In a finite system, when the rooting site $i$ lies sufficiently close to the boundary, part of this hop-distance ball intersects the exterior of the domain.
Consequently, loops that would extend beyond the boundary are excluded, and not all length-$m$ loops can be realized.
This incomplete sampling of loops near the boundary is the physical origin of finite-size corrections to spectral moments.

To formalize this, we decompose the lattice domain $\Omega$ into a bulk region $\Omega_{\mathrm b}$ and a boundary (edge) shell $\Omega_{\mathrm e}$ (of thickness $\sim m/2$ in hop distance), as illustrated in Fig.~\ref{fig:loop_counting}b.
For a fixed moment order $m$, the fully sampled sum of weights of all length-$m$ loops rooted at any bulk site $i\in \Omega_{\mathrm b}$ is identical; we denote it by $w_m$.
Defining the per-site moment
$\mathcal{M}_m := \frac{1}{N}\Tr H^m,$
one has $\mathcal{M}_m \to w_m$ as $L\to\infty$, consistent with the universal spectral moment theorem~\cite{Cheng2024UniversalSpectralMoment}.

Next, we derive the finite-size scaling law with $L$. For a boundary-shell site $i\in \Omega_{\mathrm e}$, some loops are missing because they would exit $\Omega$.
We define the missing weight at site $i$ as
$
\delta_m(i) := w_m - \sum_{|O_i|=m}w(O_i).
$
Then the relative error between $\mathcal{M}_m$ and its thermodynamic value $w_m$ can be written as
\begin{equation}\label{eq:estim}
r(m,L):=\frac{\abs{\mathcal{M}_m-w_m}}{w_m}
= \frac{\left|\sum_{i\in \Omega_{\mathrm e}} \delta_m(i)\right|}{N\,w_m}
= \frac{|\Omega_{\mathrm e}|}{N}\cdot\frac{\bar{\delta}_m}{w_m},
\end{equation}
where $\bar{\delta}_m$ is the average of $\delta_m(i)$ over $i\in \Omega_{\mathrm e}$.
For a $d$-dimensional system of linear size $L$, one has $N\sim L^d$ while the boundary area scales as $|\partial\Omega|\sim L^{d-1}$.
The boundary shell $\Omega_{\mathrm e}$ has a thickness of order $m/2$ (in hop distance), hence $|\Omega_{\mathrm e}|/N\sim m/L$ in Eq.~\eqref{eq:estim}.
Moreover, the term $\bar{\delta}_m/\omega_m$ is a non-extensive quantity which is scale independent.
Therefore, $r(m,L)$ is inversely proportional to the system length $L$, and the finite-size scaling law can be written as
\begin{equation}\label{eq:scaling}
    r(m,L)\propto \frac{f(m)}{L}
\end{equation}
where $f(m)$ is a monotonically increasing function with respect to $m$.
Intuitively, the higher the moment order $m$ is, the larger the relative deviation $r(m,L)$ will be due to a thicker boundary shell $\Omega_{\text{b}}$ causing more leakage.
The concrete form of $f(m)$ is non-universal that depends on the detailed hopping structure.
However, if the background ground potential $H_{ii}$ satisfies $|H_{ii}|\gg|H_{i\neq j}|$ (relevant to the realistic setup where the resonant frequency of cavities is much larger than the inter-cavity coupling), we can show that $f(m)\propto m^2$ (see Supplementary Materials Sec.~S1).

\subsection{Observation of spectral moments and finite-size scaling in 1D lattices}

\begin{figure}[p]
    \centering
    \phantomsection 
    \includegraphics[width=0.99\textwidth]{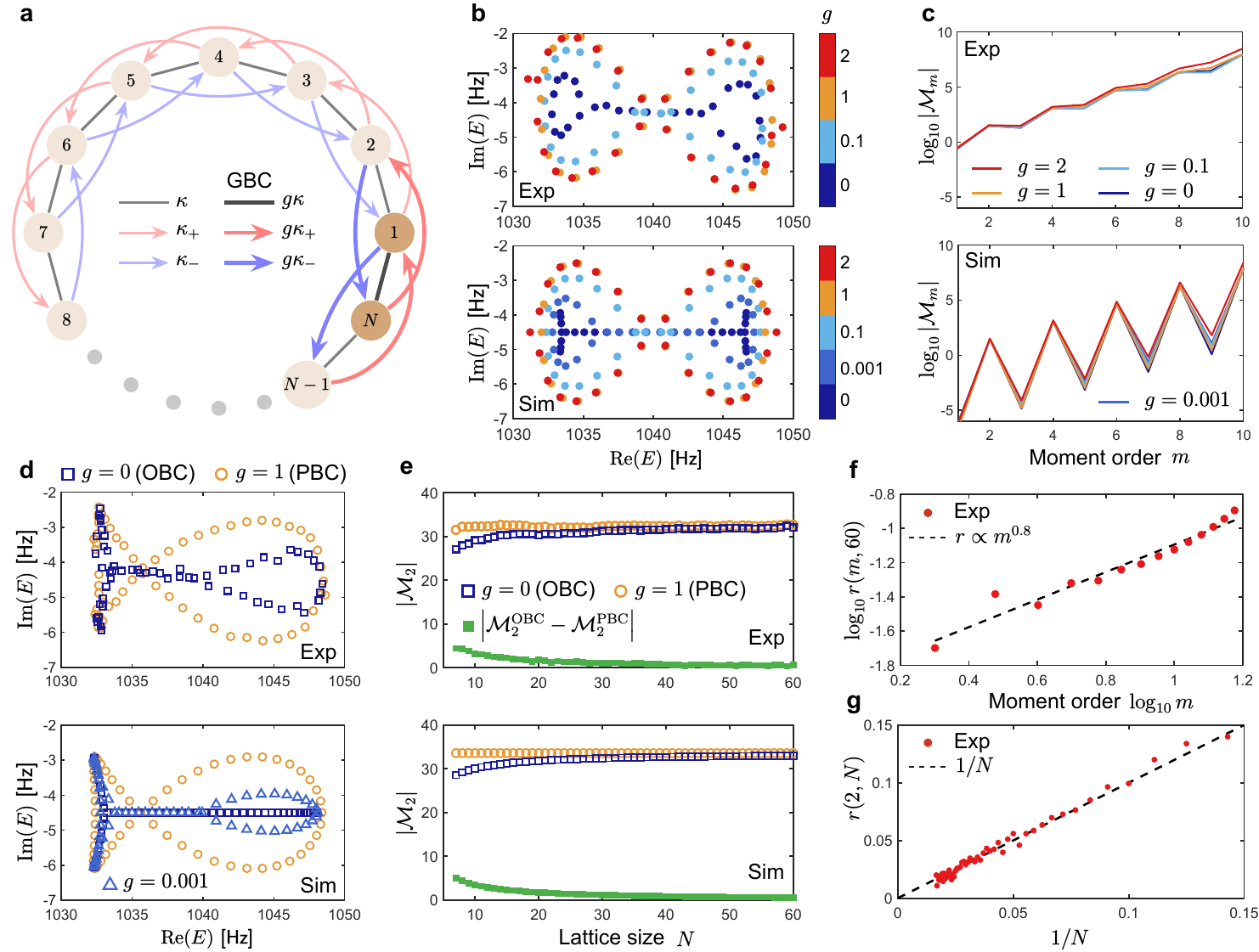}
    \caption{
        \textbf{Observation of spectral moments and finite-size scaling in a 1D non-Hermitian lattice.}
        \textbf{a,} Tight-binding schematic of the 1D non-Hermitian lattice. The Bloch Hamiltonian is
        $H(k)=\kappa_{+}\mathrm{e}^{-2\mathrm{i}k}+\kappa_{-}\mathrm{e}^{2\mathrm{i}k}+2\kappa\cos k+\omega_{0}$.
        A generalized boundary condition (GBC) continuously interpolates between open boundary condition (OBC) and periodic boundary condition (PBC) via
        $H_{\mathrm{GBC}}=H_{\mathrm{OBC}}+g\!\left(H_{\mathrm{PBC}}-H_{\mathrm{OBC}}\right)$.
        Boundary sites (sites $1$ and $N$) are highlighted in dark brown.
        \textbf{b,} Measured (top) and simulated (bottom) complex spectra under GBCs with different boundary coupling strengths $g$ (color coded). The experiment uses $g=0,\,0.1,\,1,\,2$, while the simulation additionally includes a near-OBC case $g=0.001$.
        Parameters are $\kappa_{+}=1\,\mathrm{Hz}$, $\kappa_{-}=-1\,\mathrm{Hz}$, $\kappa=4\,\mathrm{Hz}$, $\omega_{0}=1038\,\mathrm{Hz}-4.5\mathrm{i}\,\mathrm{Hz}$, and lattice size $N=30$.
        \textbf{c,} Spectral moments up to the 10th order, $\log_{10}|\mathcal{M}_{m}|$, obtained from the spectra in \textbf{b} for different $g$, showing that the moments are much less sensitive to boundary conditions than the spectra.
        \textbf{d,} Measured (top) and simulated (bottom) spectra under OBC ($g=0$) and PBC ($g=1$) for lattice size $N=60$. A simulated spectrum for $g=0.001$ is also shown to represent a near-OBC case.
        \textbf{e,} Second-order spectral moments $|\mathcal{M}_{2}|$ under OBC and PBC, together with their absolute difference $\big|\mathcal{M}^{\mathrm{OBC}}_{2}-\mathcal{M}^{\mathrm{PBC}}_{2}\big|$, as functions of lattice size $N$ (experiment, top; simulation, bottom).
        \textbf{f,} Relative deviation $r(m,60)$ extracted from the experimental OBC/PBC moments at $N=60$ (from \textbf{d}), plotted as a function of moment order $m$ on log--log scales. The dashed line is an empirical power-law fit, $r\propto m^{0.8}$, showing that the finite-size deviation increases with moment order over the experimentally accessible range.
    \textbf{g,} Relative deviation $r(2,N)$ extracted from the experimental second-order moments (from \textbf{e}) plotted against $1/N$. 
        The dashed line shows the $1/N$ scaling, confirming the finite-size law $r(m,L)\propto m^{2}/L$ (with $L=N$ for the 1D chain).
        In \textbf{d--g}, parameters are $\kappa_{+}=2\,\mathrm{Hz}$, $\kappa_{-}=0.4\,\mathrm{Hz}$, $\kappa=4\,\mathrm{Hz}$, and $\omega_{0}=1038\,\mathrm{Hz}-4.5\mathrm{i}\,\mathrm{Hz}$.}
    \label{fig:1D}
    \addcontentsline{toc}{subsection}{Figure \thefigure}
\end{figure}

We first experimentally validate the universality of spectral moments and their finite-size scaling using a one-dimensional (1D) active acoustic array \cite{Zhong2025HigherorderSkinEffect, Zhong2025ExperimentallyProbingNonHermitian, Zhong2026ObservationDislocationBound,Zhong2026ObservationErraticNonHermitian}.
The system implements a non-reciprocal tight-binding model (Fig.~\ref{fig:1D}a), in which the non-Hermitian hopping amplitudes $\kappa_+$ and $\kappa_-$ are independently controlled via electro-acoustic feedback loops \cite{Zhang2021AcousticNonHermitianSkin, Zhong2025ExperimentallyProbingNonHermitian, Wu2025ObservationDislocationNonHermitian}.
See Methods for details of the experimental setup and measurement techniques.
To systematically probe boundary dependence, we introduce a generalized boundary condition (GBC) parameterized by a boundary coupling strength $g$, which continuously interpolates between the open boundary condition (OBC, $g=0$) and the periodic boundary condition (PBC, $g=1$).
For each $g$, we measure the full complex spectrum using the Green's-function-based spectroscopy \cite{Zhong2025ExperimentallyProbingNonHermitian}, and then extract the spectral (central) moments
\begin{equation}
\mathcal{M}_m = \frac{1}{N}\sum_{j=1}^{N}(E_j-\omega_0)^m ,
\end{equation}
by summing over the measured eigenvalues $E_j$.
Here, subtracting $\omega_0$ removes the trivial on-site offset.
Without the subtraction, we show a universal scaling law  with respect to $m$ in Supplementary Materials Sec.~S1. 

Figure~\ref{fig:1D}b shows the measured and simulated spectra for a lattice of size $N=30$ under representative GBCs.
The experimental spectra are in great agreement with simulations for $g\ge 0.1$.
For the nominal OBC case ($g=0$), the measured spectrum is better reproduced by a simulation with a small residual boundary coupling ($g=0.001$), reflecting the extreme spectral sensitivity of NHSE systems to weak boundary perturbations \cite{Budich2020NonHermitianTopologicalSensors}.
As $g$ increases, the spectrum reshapes dramatically, evolving from a near arc to complex loops with substantially different geometries.
In stark contrast, the spectral moments remain nearly invariant.
Figure~\ref{fig:1D}c plots $\log_{10}|\mathcal{M}_m|$ up to the 10th order for different $g$ in both experiment and simulation.
Within each dataset, the moment curves for different boundary couplings nearly collapse onto a single trajectory, demonstrating that $\mathcal{M}_m$ is much less sensitive to boundary variations than the spectra themselves.
The absolute experimental and simulated moment values do not coincide exactly, primarily because high-order moments amplify experimental imperfections, such as small variations in onsite resonances, losses and hopping amplitudes.

We next quantify the finite-size deviation predicted by the loop-counting theory.
To this end, we compare OBC and PBC moments over lattice sizes $N=7,8,9,\dots,60$.
Figure~\ref{fig:1D}d presents the measured and simulated spectra for a representative case with $N=60$ under OBC and PBC.
Consistent with the $N=30$ case, the measured OBC spectrum is well captured by a near-OBC simulation with $g=0.001$.
Figure~\ref{fig:1D}e then shows the second-order moments $|\mathcal{M}_2|$ under OBC and PBC, together with their absolute difference $|\mathcal{M}_2^{\mathrm{OBC}}-\mathcal{M}_2^{\mathrm{PBC}}|$, as functions of $N$ (experiment and simulation).
The OBC and PBC moments converge systematically with increasing $N$, while their difference decreases toward zero.

To further quantify the finite-size deviation, we examine separately its dependence on the moment order and on the system size.
At fixed size $N=60$, Fig.~\ref{fig:1D}f plots the relative deviation $r(m,60)$ extracted from the experimental OBC/PBC moments as a function of $m$ on log--log scales.
The data show a clear monotonic increase with moment order and can be described over the experimentally accessible range by an approximate power law, $r(m,60)\propto m^{0.8}$.
This indicates that higher-order moments are more susceptible to finite-size effects, even though the accessible exponent is weaker than the asymptotic quadratic scaling predicted by the loop-counting theory.

At fixed moment order $m=2$, Fig.~\ref{fig:1D}g plots $r(2,N)$ as a function of $1/N$ for different lattice sizes.
The experimental data follow the dashed $1/N$ line closely, showing that the finite-size deviation is systematically suppressed as the system size increases.
Together, Figs.~\ref{fig:1D}f and \ref{fig:1D}g demonstrate that the deviation grows with moment order and decreases with lattice size, in qualitative agreement with the general finite-size picture developed above.

\subsection{Observation of spectral moments in higher-dimensional lattices}

\begin{figure}[p]
    \centering
    \phantomsection 
    \includegraphics[width=0.99\textwidth]{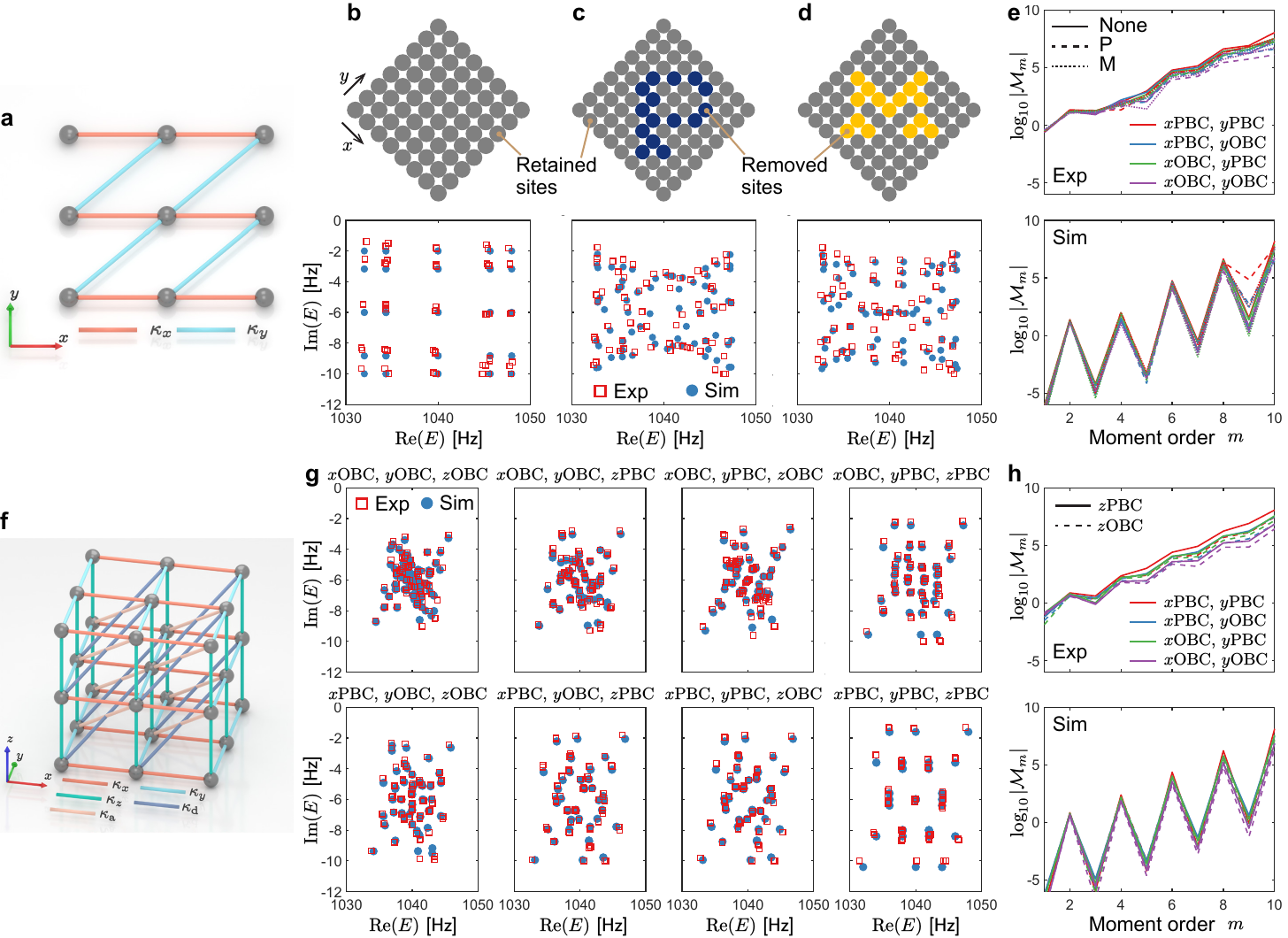}
    \caption{
        \textbf{Observation of spectral moments in higher-dimensional lattices.}
        \textbf{a}, Tight-binding schematic of the 2D reciprocal non-Hermitian lattice.
        The Bloch Hamiltonian is $H(k_x,k_y) = 2\kappa_x \cos k_x + 2\kappa_y \cos(k_x+k_y)+ \omega_0$.
        Parameters are: $\kappa_x = 2\mathrm{i\,Hz}, \kappa_y = 4\,\mathrm{Hz}$, and $\omega_0 = 1040\,\mathrm{Hz}-6\mathrm{i\,Hz}$.
        \textbf{b--d}, The top panel shows three different real-space implementations of \textbf{a} under PBCs in both $x$ and $y$ directions: \textbf{b} a regular $8\times 8 $ square lattice, and \textbf{c, d} $9\times 9$ square lattices with a few removed sites (to form an irregular boundary) to show the letter \textbf{c} `P' and \textbf{d} `M'.
        The bottom panel in \textbf{b--d} shows the measured (red squares) and simulated (blue dots) energy spectra.
        \textbf{e}, Spectral moments up to the 10th order obtained from experiments (top) and simulations (bottom) using data in \textbf{b--d}.
        \textbf{f}, Tight-binding schematic of the 3D reciprocal non-Hermitian lattice.
        The Bloch Hamiltonian is $H(k_x,k_y,k_z) = \omega_0 + 2\kappa_x \cos k_x + 2\kappa_y \cos k_y + 2\kappa_z \cos k_z + \kappa_\mathrm{d}[\mathrm{e}^{\mathrm{i}(k_x+k_y+k_z)} + \mathrm{e}^{-\mathrm{i}(k_x+k_y+k_z)}] + \kappa_\mathrm{a} [\mathrm{e}^{\mathrm{i}(k_x-k_y+k_z) + \mathrm{i}(k_x+k_y-k_z)}]$.
        Parameters are: $\kappa_x = \mathrm{i\,Hz}, \kappa_y = 1\,\mathrm{Hz}, \kappa_z = 1\,\mathrm{i\,Hz}, \kappa_\mathrm{d} = 2\,\mathrm{Hz}, \kappa_\mathrm{a} = 1.2\,\mathrm{Hz}$, and $\omega_0 = 1040\,\mathrm{Hz}-6\mathrm{i\,Hz}$.
        The lattice size is $4\times 4\times 4$.
        \textbf{g}, Measured (red squares) and simulated (blue dots) energy spectra under different boundary conditions.
        \textbf{h}, Spectral moments up to the 10th order obtained from experiments (top) and simulations (bottom) using data in \textbf{g}.
    }
    \label{fig:higherD}
    \addcontentsline{toc}{subsection}{Figure \thefigure}
\end{figure}

To test the dimensional robustness of spectral moments and their insensitivity to boundary reshaping beyond 1D, we extend our measurements to both 2D and 3D non-Hermitian acoustic lattices.
We first implement a 2D reciprocal non-Hermitian lattice (Fig.~\ref{fig:higherD}a) and realize three distinct real-space geometries: a regular square array and two irregular domains shaped as the letters ``P'' and ``M'', obtained by selectively removing sites from a larger square lattice (Figs.~\ref{fig:higherD}b--\ref{fig:higherD}d, top).
These geometries are chosen to deliberately induce strong boundary reshaping while preserving the same bulk hopping pattern.

The measured spectra for the three 2D geometries (Figs.~\ref{fig:higherD}b--\ref{fig:higherD}d, bottom) agree well with numerical simulations and are clearly distinct from one another in the complex plane, demonstrating pronounced geometry-dependent spectral response in higher-dimensional non-Hermitian lattices \cite{Zhang2022UniversalNonHermitianSkin, Wang2023ExperimentalRealizationGeometrydependent, Zhou2023ObservationGeometrydependentSkin, Wan2023ObservationGeometrydependentSkin,ultrasensitivity2025, Zhong2025ExperimentallyProbingNonHermitian}.
We then compute spectral moments for all measured 2D configurations, including different boundary-condition choices along the $x$ and $y$ directions (PBC/OBC combinations) and the three real-space geometries (``None'', ``P'', and ``M'').
As shown in Fig.~\ref{fig:higherD}e, although the underlying spectra vary substantially, the moments up to the 10th order collapse onto nearly the same curves in both experiment and simulation, in agreement with the bulk-theory prediction.
This provides a higher-dimensional confirmation that spectral moments are far more boundary robust than the spectra themselves.

We next implement a 3D reciprocal non-Hermitian lattice (Fig.~\ref{fig:higherD}f) on a $4\times4\times4$ cubic array with programmable couplings.
This platform enables independent boundary control along all three directions, allowing us to probe PBC/OBC and mixed boundary conditions in $(x,y,z)$.
Figure~\ref{fig:higherD}g shows the measured and simulated complex spectra for representative boundary-condition combinations.
As in 2D, the spectra exhibit strong boundary dependence and substantial reshaping across different boundary configurations, while the experiment remains in good agreement with simulation.

Despite these dramatic spectral changes, the corresponding spectral moments remain nearly invariant.
Figure~\ref{fig:higherD}h summarizes the moments extracted from the 3D spectra for different $(x,y,z)$ boundary-condition combinations, showing a strong collapse of the moment curves in both experiment and simulation.
Together, the 2D and 3D results establish that the universality of spectral moments persists across dimensions and under highly nontrivial boundary reshaping, with the remaining discrepancies attributable to finite-size corrections and experimental uncertainty.

\subsection{Observation of the dispersive-to-proliferative dynamical phase transition}

\begin{figure}[p]
    \centering
    \phantomsection 
    \includegraphics[width=0.99\textwidth]{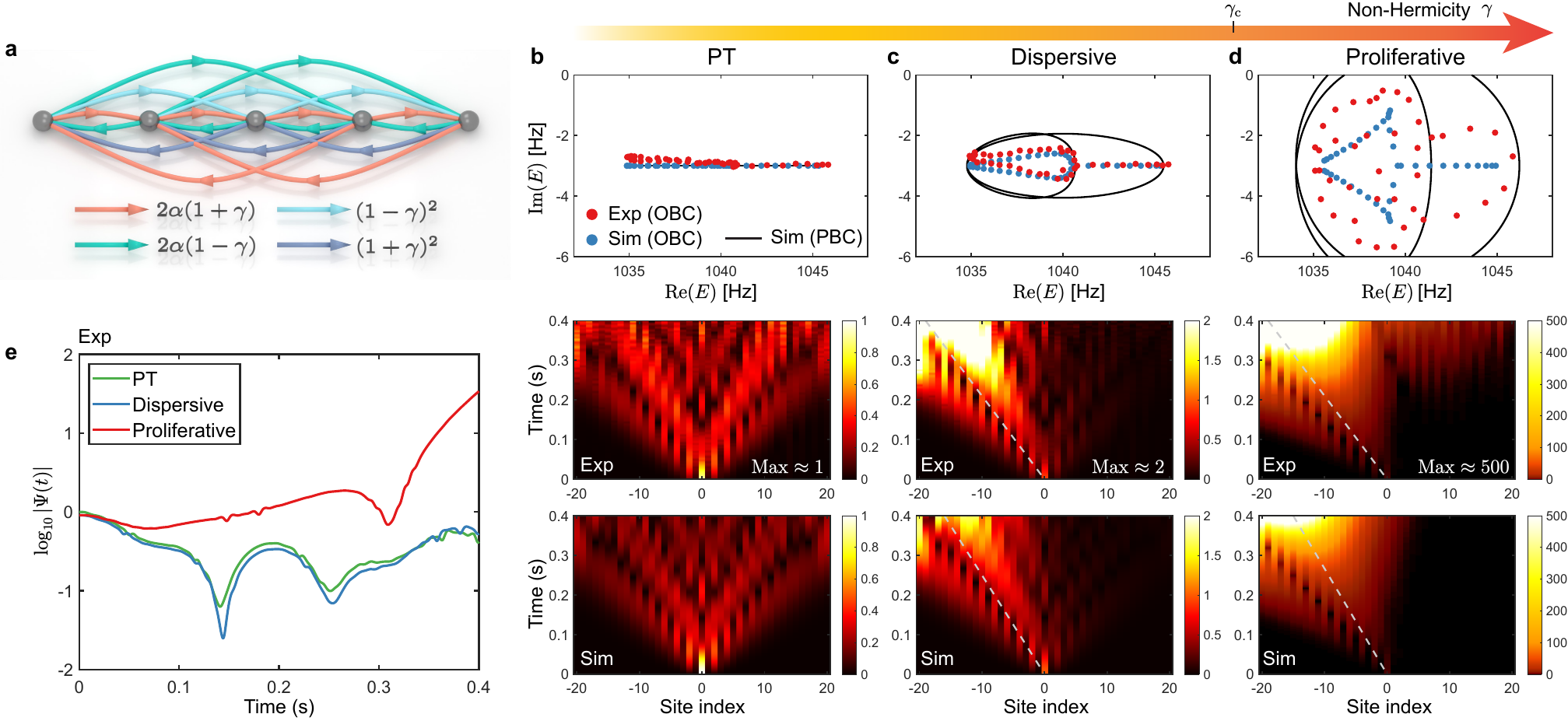}
    \caption{
        \textbf{Observation of the dynamic dispersive-to-proliferative phase transition.}
        \textbf{a}, Tight-binding schematic of the 1D non-Hermitian PT-symmetric lattice.
        The Bloch Hamiltonian is $H(k) = 2\alpha(1-\gamma)\ee^{-3\ii k} + (1-\gamma)^2 \ee^{-2\ii k} + 2\alpha(1+\gamma) \ee^{-\ii k} + 2\alpha(1-\gamma) \ee^{\ii k} + (1+\gamma)^2 \ee^{2\ii k} + 2\alpha(1+\gamma)\ee^{3\ii k}$.
        \textbf{b--d}, Results for the three representative regimes: \textbf{b} $\mathcal{PT}$-symmetric ($\gamma=0$), \textbf{c} dispersive ($\gamma=0.13$), and \textbf{d} proliferative ($\gamma=0.45$).
        In each of \textbf{b--d}, the top panel shows the measured OBC spectra (red dots), the simulated OBC spectra (blue dots), and the theoretical PBC spectra (black curves).
        The middle and bottom panels show the measured and simulated spatiotemporal evolution of the wave-function amplitude for an initial excitation at site 0, respectively.
        Dashed lines indicate the dominant propagation direction.
        For $\gamma>0$, the left-right hopping imbalance produces directional, asymmetric propagation in both the dispersive and proliferative regimes.
        The key distinction is the amplitude scale: in the proliferative regime (\textbf{d}), the loss-compensated amplitude grows by more than two orders of magnitude within the experimental time window, as reflected by the much larger colorbar scale in \textbf{d} compared with \textbf{c}.
        \textbf{e}, Time evolution of the wave-function amplitude at the source site (site 0) for the three regimes in \textbf{b--d}, highlighting decay/dispersion in the $\mathcal{PT}$-symmetric and dispersive regimes and rapid growth in the proliferative regime.
    }
    \label{fig:dyn}
    \addcontentsline{toc}{subsection}{Figure \thefigure}
\end{figure}

Having established spectral moments as boundary-robust bulk observables, we next probe their dynamical implications using the same unified spectral-and-dynamical acoustic platform.
We focus on a 1D $\mathcal{PT}$-symmetric non-Hermitian lattice (Fig.~\ref{fig:dyn}a), in which the non-Hermiticity is tuned by a control parameter $\gamma$.
According to the spectral-moment framework, short-time local bulk dynamics are governed by the bulk moment structure, and therefore need not follow the strong boundary sensitivity of the OBC spectrum~\cite{Cheng2024UniversalSpectralMoment}.
This predicts a counterintuitive possibility: local bulk dynamics can remain dispersive even when the OBC spectrum is already complex.

To test this prediction, we systematically vary $\gamma$ and, for each value, measure both the complex OBC spectrum and the time-domain evolution of a wave packet initially excited at a bulk site (site 0).
To isolate the intrinsic non-Hermitian dynamics from the uniform background dissipation of the acoustic platform, we compensate the measured signals by a virtual gain factor $\exp[-2\pi \,\mathrm{Im}(\omega_0)t]$, where $-\mathrm{Im}(\omega_0)\approx 3~\mathrm{Hz}$.
Figures~\ref{fig:dyn}b--\ref{fig:dyn}d summarize the results for three representative regimes.
In each case, the top panel compares measured OBC spectra, simulated OBC spectra, and the theoretical PBC spectrum, while the middle and bottom panels show the measured and simulated spatiotemporal wave-packet dynamics, respectively.
The experimental data are in good agreement with simulations in both spectral and dynamical observables.


At $\gamma=0$, the system is in the $\mathcal{PT}$-symmetric regime (Fig.~\ref{fig:dyn}b).
The spectrum is purely real relative to the common loss offset, and the wave packet exhibits stable spreading dynamics without amplification.
Because the left- and right-going hopping amplitudes are balanced at $\gamma=0$, the measured propagation is approximately symmetric about the initially excited site.
When $\gamma$ is increased to $\gamma=0.13$, below the theoretical critical value $\gamma_{\mathrm c}\approx 0.155$ (see Supplementary Materials), the system enters the \emph{dispersive} regime (Fig.~\ref{fig:dyn}c).
The nonzero $\gamma$ introduces a left-right hopping imbalance, leading to asymmetric, directionally biased wave-packet propagation.
Importantly, the measured OBC spectrum is already complex and exhibits loop-like spectral structures.
Nevertheless, the loss-compensated wave amplitude remains bounded and dispersive: although the wave packet drifts asymmetrically, its amplitude scale stays of order unity within the experimental time window.
This demonstrates that complex, boundary-sensitive OBC spectra, even when referenced to the uniform-loss-compensated frame, do not by themselves determine the local bulk dynamical phase.


Upon further increasing the non-Hermiticity to $\gamma=0.45>\gamma_{\mathrm c}$, the system crosses into the \emph{proliferative} regime (Fig.~\ref{fig:dyn}d).
The wave packet remains directionally biased because the hopping imbalance persists, but its amplitude behavior is qualitatively different from the dispersive case.
After compensating the uniform background loss, the wave amplitude rapidly increases and reaches values more than two orders of magnitude larger than in the dispersive regime within the same experimental time window, as indicated by the much larger colorbar scale in Fig.~\ref{fig:dyn}d.
Figure~\ref{fig:dyn}e further highlights this distinction by comparing the source-site amplitude, which remains bounded or decays in the $\mathcal{PT}$-symmetric and dispersive regimes but grows rapidly in the proliferative regime.

These results provide an experimental demonstration of the dispersive-to-proliferative transition as an intrinsic bulk dynamical phenomenon governed by bulk moment structure.
More broadly, they show that dynamical stability in finite non-Hermitian lattices can be decoupled from direct spectral expectations based solely on boundary-sensitive OBC spectra in the presence of a uniform background loss, thereby extending the significance of spectral-moment invariance from static spectroscopy to time-domain wave control.

\section{Conclusions}
In conclusion, we present the first experimental validation of the universal spectral moment theorem and establish spectral moments as experimentally accessible, boundary-robust descriptors of bulk non-Hermitian lattices. Using a unified spectral-and-dynamical acoustic platform, we bridge the gap between thermodynamic-limit theory and finite-size experimental reality by deriving and verifying universal finite-size scaling laws across 1D, 2D, and 3D systems. These results show that, even when spectra are drastically reshaped by boundary geometry through the non-Hermitian skin effect, spectral moments remain a reliable probe of intrinsic bulk physics, with quantitatively controlled finite-size corrections.

Beyond spectroscopy, our time-domain measurements uncover the predicted dispersive-to-proliferative transition and reveal a counterintuitive dynamical regime in which local bulk dynamics remain dispersive despite strongly boundary-sensitive complex OBC spectra. This demonstrates that local bulk dynamical stability need not be inferred from boundary-shaped spectra alone, and extends the significance of spectral-moment invariance from static spectral characterization to dynamical phase diagnostics. More broadly, our work provides a practical route for extracting and controlling intrinsic bulk behavior in finite non-Hermitian wave systems, and opens opportunities for boundary-robust functionalities in non-Hermitian metamaterials and wave-based devices.

\section{Acknowledgment}
Y. J. thanks NSF CMMI awards 2039463 and 195122.
C. S, N. C., K. Z., and K. S. acknowledge the support by the Office of Naval Research (Grant No. MURI N00014-20-1-2479) and the National Science Foundation through the Materials Research Science and Engineering Center at the University of Michigan (Award No. DMR-2309029).

\section{Author contributions}
J.-X. Z., K. S., and Y. J. conceived the project.
C. S, N. C., and K. Z. performed theoretical analysis and numerical simulations.
J.-X. Z. designed and performed the experiments with assistance from J. W. K..
J.-X. Z., C. S., K. S., and Y. J. wrote the paper.
All authors reviewed the paper.
K. S. and Y. J. supervised the project.

\section{Competing interests}
The authors declare no competing interests.

\clearpage
\printbibliography
\addcontentsline{toc}{section}{References}

\clearpage 
\section{Methods}

\subsection{Implementation of non-Hermitian acoustic lattices}

The non-Hermitian acoustic lattices consists of multiple 3D-printed acoustic cavities (see Supplementary Materials for dimensions), each representing a site in the non-Hermitian lattice and operating near its first dipole resonance around 1040\,Hz.
The tuning of onsite potentials and hoppings is implemented using active components, specifically loudspeaker-microphone (pump-probe) pairs, an audio amplifier (LM386, Texas Instruments), and an phase shifter integrated in a customized controller.
The loudspeaker and microphone are positioned at the bottom of each cavity. 
The microphone captures the acoustic pressure signal, which is processed by the amplifier and phase shifter to adjust its amplitude and phase before being emitted by the loudspeaker in the connected cavity (see \cite{Zhong2025ExperimentallyProbingNonHermitian} for details). 

\subsection{Spectral reconstruction via Green's function measurements}

For a chain of $N$ sites, we measure the steady-state complex response at all probe sites $i$ for each pump site $j$ under monochromatic excitation at frequency $\omega$.
This yields the full response matrix $G_{ij}(\omega)$ over a discrete set of drive frequencies spanning the band of interest.
The magnitude and phase are obtained by referencing the microphone signals to the excitation signal, allowing coherent extraction of $G_{ij}(\omega)$.
The full matrix acquisition over all pump-probe pairs yields $N^2$ complex transfer functions per frequency point \cite{Zhong2025ExperimentallyProbingNonHermitian}.

In the effective tight-binding description, the Green's function satisfies
\begin{equation}
G(\omega) = (\omega - H)^{-1},
\end{equation}
so the eigenvectors of $G(\omega)$ coincide with those of $H$, and its eigenvalues follow the single-pole form 
\begin{equation}
\lambda_m(\omega)=\frac{1}{\omega-E_m}.
\label{eq:pole_fit}
\end{equation}
up to a complex prefactor set by the measurement normalization.
We therefore diagonalize $G(\omega)$ for each sampled $\omega$, track each eigenvalue branch $\lambda_m(\omega)$ across frequency, and fit to Eq.~\eqref{eq:pole_fit} to extract $E_m$.
This procedure yields both $\Re(E_m)$ and $\Im(E_m)$, allowing direct access to the complex spectrum.

\subsection{Time-resolved propagation measurements}

In the dynamical measurements, we excite the lattice with a Gaussian-modulated tone burst $s(t)$ centered at frequency $\Re (\omega_0)$ 
\begin{equation}
    s(t)=
    \exp\qty[{-\frac{(t-t_0)^2}{2\sigma^2} }-{i} \Re\qty(\omega_{0})t] 
    .
    \label{eq:gaussian_pulse}
\end{equation}
Here, $t_0 $ is the pulse center time, and $\omega_0$ is the complex resonant frequency of a single cavity with the imaginary part accounting for intrinsic losses. 
See Supplementary Materials for details on the temporal and spectral characteristics of the excitation signals.
The pulse width $\sigma$ is chosen to satisfy two experimental constraints:
(i) $\sigma$ cannot be too small, otherwise the injected energy becomes insufficient and the propagating wave packet rapidly decays into the noise floor;
(ii) $\sigma$ cannot be too large, otherwise the excitation spectrum becomes too narrow to cover the relevant bandwidth of the lattice spectrum.
The time-resolved signals are simultaneously recorded at all sites using synchronized data acquisition modules. 
Importantly, the envelope of these time-domain measurements is extracted to represent the wavefunctions $\psi_n(t)$.

In experiments, the carrier frequency is set to $\Re(\omega_0)=1040~\mathrm{Hz}$ (near the band center), and the envelope width is chosen as $\sigma =10\,\mathrm{ms}$.
The driving signal is generated by a multi-channel data-acquisition card (PCIe-6353, National Instruments) and applied to
the on-site loudspeaker (CMS-15118D-L100, CUI Devices) through an audio amplifier (LM386, Texas Instruments).
The time-domain waveform at each site is recorded simultaneously using MEMS microphones (BOB-19389, SparkFun Electronics) at a
sampling rate of $16~\mathrm{kHz}$, with all channels acquired synchronously.

The recorded pressure signals are digitally bandpass-filtered around the carrier frequency (1010--1070~Hz).
The envelope of the signal is taken as $\abs{\psi_n(t)}$.
At each time $t$, we normalize $\psi_n(t)$ by its instaneous norm, so that $\sum_n \abs{\psi_n(t)}^2=1$ for all $t$.


\clearpage
\section{Figures}
\vfill
\vfill

\clearpage

\clearpage

\clearpage

\end{document}